\documentclass[12pt,preprint]{aastex}

\newcommand{\HCOP}{HCO$^+$}
\newcommand{\HICOP}{H$^{13}$CO$^+$}

\newcommand{\jone}{J=1$\rightarrow$0}
\newcommand{\jtwo}{J=2$\rightarrow$1}
\newcommand{\jthree}{J=3$\rightarrow$2}
\newcommand{\jfour}{J=4$\rightarrow$3}

\newcommand{\jseven}{J=7$\rightarrow$6}

\newcommand{\Kkms}{K$\cdot$km\,s$^{-1}$}
\newcommand{\kms}{km\,s$^{-1}$}
\newcommand{\Msun}{M$_\odot$}
\newcommand{\Lsun}{L$_\odot$}
\newcommand{\Rsun}{R$_\odot$}

\slugcomment{To appear in ApJ}

\shorttitle{Infall towards Serperns SMM4}
\shortauthors{Narayanan et al.}

\begin{document}


\title{Detection of Infall Signatures Towards Serpens SMM4}

\author{G. Narayanan} 
\affil{Five College Radio Astronomy Observatory, University of Massachusetts,
    Amherst, MA 01003}
\email{gopal@astro.umass.edu}

\author{G. Moriarty-Schieven}
\affil{Joint Astronomy Centre, 660 North Aohoku Place, University
Park, HI 96720}

\author{C. K. Walker}
\affil{Steward Observatory, University of Arizona, Tucson, AZ 85721}

\and

\author{H. M. Butner}
\affil{Submillimeter Telescope Observatory, University of Arizona,
Tucson, AZ 85721}

\begin{abstract}
We present the detection of kinematic infall signatures towards the
Class 0 protostellar system SMM4 in the Serpens cloud core. We have
observed the dense molecular gas towards the embedded source using
millimeter and submillimeter line transitions of density sensitive
molecular tracers. 
High signal-to-noise ratio maps obtained in \HCOP\ \jone, \jthree\ and
\jfour, and CS \jtwo\ show the blue-bulge infall signature. The
blue-bulge infall signature can be observed in the centroid velocity
maps of protostellar objects when infall dominates over rotation. The
line profiles of \HCOP\ and CS exhibit the characteristic blue
asymmetric line profile signature consistent with infall. In addition,
\HCOP\ and CS optical depth profiles obtained using isotopic
observations show a red asymmetry also consistent with an infall
interpretation. Using three-dimensional radiative transfer models
based on the rotating, collapse model of Terebey, Shu and Cassen, we
derive infall parameters of the source. To determine the direction and
orientation of molecular outflows in the larger Serpens cluster,
wide-field mapping of CO \jone\ emission was also performed.

\end{abstract}


\section{Introduction}

The study of star formation is currently at a fascinating stage. With
the advent of large millimeter and submillimeter single-dish
telescopes, and interferometers, the earliest stages of star formation
have been under considerable observational scrutiny. The youngest
protostars, also known as Class 0 sources (of age a few $\times 10^4$
yrs), are so completely embedded in their dust and gas cocoons that
they are only detectable longward of mid-infrared wavelengths
\citep*{and93,bar94}. A typical Class 0 protostar consists of a central
accreting source (that may be surrounded by a dynamic accretion disk),
an extended, massive, rotating infall envelope, and a vigorous bipolar
molecular outflow. 

While collapse is one of the most important stages of star formation,
direct observation of protostellar collapse is rendered difficult due
to the kinematic confusion in star forming regions. Millimeter and
submillimeter line profiles are not just affected by collapse motions.
They are also affected by motions such as rotation, outflow and
turbulence that are characteristic of the earliest stages of star
formation. Much effort has been invested in finding a distinctive
signature of infall. The most used infall signature is the blue
asymmetric line profile that is seen in optically thick molecular
transitions in the presence of a centrally peaked excitation
temperature gradient \citep{wal86,zho92,zho95}. A {\em mapping}
signature based on centroid velocity maps (dubbed the ``blue-bulge''
signature of infall) was proposed and subsequently detected towards
several Class 0 sources \citep*{wnb94, nar97, nar98, nwb98}. As long
as the emission remains optically thick, the blue-bulge signature is
found to be relatively insensitive to both variations in molecular
abundance and source inclination \citep{nar98}. The blue-bulge
signature naturally accounts for infall and rotation, and since it
occurs along the equatorial direction of the protostar, is expected to
be less affected by outflow.

The Serpens star-forming core, at a distance of 310 pc, is known to
harbor about a half-dozen Class 0 protostellar candidates
\citep*{cas93,white95,hurt96,testi98}. As part of a study of the
kinematics and evolutionary state of a larger set of YSOs
\citep{nar2001}, we performed a multi-transitional study of all the
Class 0 objects in the Serpens core. In this paper, we concentrate on SMM4,
which is the brightest submillimeter continuum object in the
southeastern section of the Serpens core \citep{cas93}, and has the
strongest overall H$_2$CO $3_{03}\rightarrow 2_{02}$ emission of all
observed sources in Serpens\citep{hurt96}. We mapped the
CS \jtwo\ transition, and three different transitions of \HCOP\  emission
at four different angular resolutions to constrain the
infall parameters of this source. We also mapped a 20\arcmin\ $\times$
10\arcmin\ region around SMM4 in CO \jone\ to determine the orientation
and effect of outflows.  In addition, we constrained the infall
parameters of SMM4 using three-dimensional collapse models based on
the \citep*[hereafter TSC]{tsc84} solutions for
protostellar collapse. In \S2 we describe our observations, and in \S3
we present our results.

\section{Observations}

A summary of all the observations obtained is shown in Table
\ref{tbl-1}. 

\subsection{FCRAO Observations}

Observations at the Five College Radio Astronomy
Observatory\footnote{FCRAO is supported in part by the National
Science Foundation under grant AST 97-25951.} (FCRAO) 14~m telescope
were performed in December, 1998 using the SEQUOIA 16-beam array
receiver \citep*{erick99}, and the FAAS backend consisting of 16
autocorrelation spectrometers. The effective resolution obtained with
each transition is summarized in Table 1. The \HCOP, CS and their
isotopic counterpart transitions were observed using the
frequency-switched mode and, after folding, third order baselines were
subtracted. $^{12}$CO and its rarer isotopes were observed using
position-switched mode, and first order baselines were
removed. Pointing and focus were checked every few hours on nearby SiO
maser sources. A 6\arcmin\ $\times$ 6\arcmin\ region centered on SMM4
($\alpha$ (1950) = $18^h27^m24.7^s$, $\delta$ (1950) = 1\arcdeg
11\arcmin 10\arcsec) was mapped with half-beam sampling in all
transitions. In the CO \jone\ transition, a larger region (20\arcmin\
$\times$ 12\arcmin) was mapped at full beam sampling.

\placetable{tbl-1}

\begin{deluxetable}{llccc}
\tablecolumns{5}
\footnotesize
\tablecaption{Observations. \label{tbl-1}}
\tablewidth{0pt}
\tablehead{
\colhead{Telescope} & \colhead{Transition}   & \colhead{Frequency} 
& \colhead{Velocity Res.} & \colhead{Beam
Size} \\
&& \colhead{(GHz)} & \colhead{(kms$^{-1}$)} & \colhead{(\arcsec)}}
\startdata
FCRAO & \HICOP\  \jone & 86.754330 & 0.067 & 62 \\
FCRAO & \HCOP\  \jone & 89.188523 & 0.066 & 60 \\
FCRAO & C$^{34}$S \jtwo & 96.412940 & 0.061 & 56\\
FCRAO & C$^{32}$S \jtwo & 97.980950 & 0.060 & 55\\
FCRAO & $^{13}$CO \jone & 110.201354 & 0.053 & 49 \\
FCRAO & CO \jone & 115.271202 & 0.203 & 47 \\
\tableline
JCMT & \HICOP\  \jthree & 260.255478 & 0.090 & 19 \\
JCMT & \HCOP\  \jthree & 267.557619 & 0.088 & 19 \\
\tableline
HHT & \HICOP\  \jthree & 260.255478 & 0.288 & 29 \\
HHT & \HCOP\  \jthree & 267.557619 & 0.288 & 28 \\
HHT & $^{13}$CO \jthree & 330.587960 & 0.227 & 23 \\
HHT & CO \jthree & 345.795999 & 0.217 & 22 \\
HHT & \HICOP\  \jfour & 346.998540 & 0.210 & 22 \\
HHT & \HCOP\  \jfour & 356.734256 & 0.210 & 21 \\
\enddata
\end{deluxetable}

\subsection{HHT Observations}

\HCOP\  and \HICOP\  \jfour\ and CO \jthree\ observations were
conducted in January 1998 with the 10~m Heinrich Hertz Telescope
(HHT)\footnote{The HHT is operated by the Submillimeter Telescope
Observatory (SMTO), and is a joint facility for the University of
Arizona's Steward Observatory and the Max-Planck-Instit\"ut fur
Radioastronomie (Bonn).}. The facility dual polarization 345 GHz SIS
receiver system was used as the frontend. The \jthree\ transitions of
\HCOP\  and \HICOP\  were observed in June 1998 with the facility 230
GHz SIS receiver.  For both sets of observations, several backend
spectrometers were used simultaneously. The available spectrometers at
the HHT were two 1 GHz wide ($\sim$ 1 MHz resolution) acousto-optic
spectrometers (AOSs), one 250 MHz wide ($\sim$ 400 kHz resolution) AOS,
and three filterbank spectrometers (with resolutions of 1 MHz, 250 kHz
and 62.5 respectively). The results presented in this paper
are only from the 250 kHz filterbank spectrometer. All observations
were conducted using position-switching. The \HCOP\  and CO transitions
were mapped in a 1\arcmin\ $\times$ 1\arcmin\ region centered on SMM4
with a grid spacing of 10\arcsec. Single spectra were obtained towards
SMM4 with the \HICOP\  transitions. 

\subsection{JCMT Observations}

\HCOP\  \jthree\ observations were also carried out at the James Clerk
Maxwell Telescope (JCMT)\footnote{The James Clerk Maxwell Telescope is
operated by the Royal Observatories on behalf of the United Kingdom
Particle Physics and Astronomy Research Council, the Netherlands
Organization for Scientific Research and the Canadian National
Research Council.} in August 1998. A 1\arcmin\ $\times$ 1\arcmin\ map
was made towards SMM4, with a grid spacing of 10\arcsec. The mapping
was performed using the on-the-fly mapping capability at the
JCMT. Position switched observations were also made in the \HICOP\ 
\jthree\ transition towards the central position of SMM4. The JCMT
observations were done using the facility A3i single channel SIS
receiver. The spectrometer backend used was the Dutch Autocorrelation
Spectrometer (DAS) configured to operate with an effective resolution
of 95 kHz and total bandwidth of 125 MHz.

\section{Analysis}

\subsection{Line Profiles}

\begin{figure*}
\epsscale{0.9}
\plotone{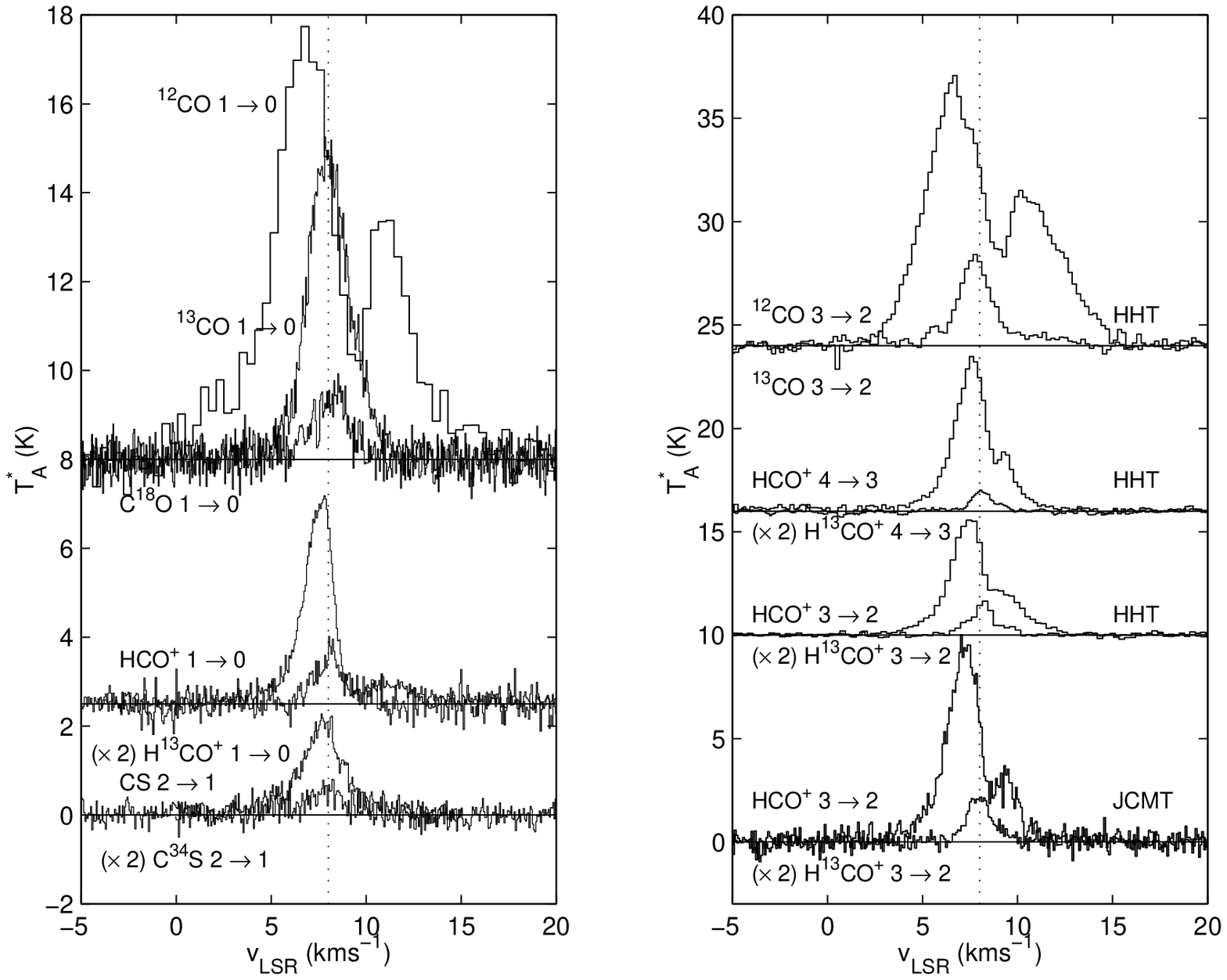}
\caption{Observed line profiles towards the central position of
SMM4. The left panel shows spectra obtained at FCRAO, while the right
panel shows the submillimeter spectra obtained with the telescope
indicated next to the spectra. The velocity extent of the displayed
spectra are $-5$ to $20$ kms$^{-1}$. The vertical dotted line
represents the systemic velocity ($8$ kms$^{-1}$) of the source. See
Table \ref{tbl-1} for the effective angular and spectral resolution
for each observed transition.\label{plotall}}
\end{figure*}

\placefigure{plotall}

In Figure \ref{plotall} we present a plot of our spectral observations
towards the central position of SMM4. The left panel shows the
millimeter lines and the right panel shows the submillimeter lines. In
Table \ref{tbl-2}, we list the centroid velocities with uncertainties
of the transitions shown in Figure \ref{plotall}. The centroid
velocities were computed over a velocity interval that corresponded to
the linewidth of the optically thin isotope.  Also listed in the table
are the beamwidths of the observations repeated from Table
\ref{tbl-1}. 

Several trends can be observed in the \HCOP\ and CS spectra towards
the central position that suggest that infall is occurring towards
SMM4. The noteworthy feature of most of the \HCOP\ and CS lines (the
main isotope) in Figure \ref{plotall} is that they show the classic
blue asymmetric line profile, the signature expected for infall. The
millimeter transitions of CS and \HCOP\ do not show identifiable
double-peaked line profiles. The self-absorption dip in the \HCOP\
\jone\ line may actually be down at the continuum level (there is some
emission redward of this dip; see Figure\ref{plotall}). While the CS
\jtwo\ line profile does not show a clear blue asymmetry, its centroid
velocity is clearly blueshifted (see Table \ref{tbl-2}). In contrast,
the optically thin isotopic spectra (the \HICOP\ and C$^{34}$S
transitions) appear more gaussian and centered on the v$_{LSR}$ of the
object, again as expected in an infall interpretation
\citep{nwb98,nar98}. 

In all cases, for molecular tracers that probe the infalling material,
the centroid velocity of the corresponding main, more optically thick
isotope would be expected to be lower (bluer) than its rarer, more
optically thin counterpart. Indeed that is the pattern seen for
corresponding isotopic pairs for \HCOP\ and CS in Table
\ref{tbl-2}. It would also be expected that higher lying submillimeter
transitions show a more pronounced blue asymmetric line profile, since
they would preferentially probe regions of greater density and hence
larger infall velocities. Indeed, the HHT \HCOP\ \jfour\ transition
shows a more pronounced blue asymmetry than the HHT \HCOP\ \jthree\
transition. However, the JCMT \HCOP\ \jthree\ transition has a bluer
centroid velocity than the HHT \HCOP\ \jfour\ transition (see Table
\ref{tbl-2}). This is probably because the JCMT \jthree\ observation
has a higher angular resolution than the HHT \jfour\ observation.  At
higher angular resolution, the same transition would probe more
embedded regions of higher infall velocity, and hence would be
expected to have more pronounced blue asymmetry. Indeed that is the
case seen towards the higher resolution JCMT HHT \HCOP\ \jthree\
spectrum as compared with the corresponding HHT spectrum.  The
corresponding centroid velocity for the JCMT observed \jthree\
transition is also lower (or bluer) compared to the HHT observed
transition (see Table \ref{tbl-2}). As expected for an infall
interpretation, the FCRAO millimeter wavelength CS and \HCOP\ line
profiles, on account of their lower angular resolution and lower
critical densities appear to show a less pronounced blue asymmetry
compared to the submillimeter transitions. 
The FCRAO CO \jone\ and the HHT CO \jthree\ transitions also show blue
asymmetric line profiles. Their respective isotopic transitions are
more gaussian in appearance and centered about the v$_{LSR}$ of the
source. Low-lying CO transitions, because of their lower critical
density are not expected to be as good tracers of infall as \HCOP\ or
CS. In addition, CO is usually seen to trace outflows in protostellar
objects. Indeed, in SMM4, as we will show below, the $^{12}$CO
transitions do trace outflow in this system. Table~\ref{tbl-2} shows
that for the submillimeter transition of CO, the main isotope has a
redder centroid velocity that the corresponding rarer isotope. This
would be consistent with CO \jthree\ tracing the underlying
outflow. 

\placetable{tbl-2}

\begin{deluxetable}{llccc}
\tablecolumns{5}
\footnotesize
\tablecaption{Centroid Velocities \label{tbl-2}}
\tablewidth{0pt}
\tablehead{
\colhead{Telescope} & \colhead{Transition}   & \colhead{v$_C$} 
& \colhead{$\sigma_{v_C}$} & \colhead{Beam
Size} \\
&& \colhead{(kms$^{-1}$)} & \colhead{(kms$^{-1}$)} & \colhead{(\arcsec)}}
\startdata
FCRAO & \HICOP\  \jone & 8.17 & 0.05  & 62 \\
FCRAO & \HCOP\  \jone & 7.66 & 0.09 & 60 \\
FCRAO & C$^{34}$S \jtwo & 7.97 & 0.01 & 56\\
FCRAO & C$^{32}$S \jtwo & 7.62 & 0.01 & 55\\
FCRAO & $^{13}$CO \jone & 8.02 & 0.01 & 49 \\
FCRAO & CO \jone & 7.77 & 0.12 & 47 \\
\tableline
JCMT & \HICOP\  \jthree & 7.97 & 0.01 & 19 \\
JCMT & \HCOP\  \jthree & 7.43 & 0.05 & 19 \\
\tableline
HHT & \HICOP\  \jthree & 8.04 & 0.01 & 29 \\
HHT & \HCOP\  \jthree & 7.63 & 0.01 & 28 \\
HHT & $^{13}$CO \jthree & 7.90 & 0.10 & 23 \\
HHT & CO \jthree & 8.06 & 0.02 & 22 \\
HHT & \HICOP\  \jfour & 8.11 & 0.02 & 22 \\
HHT & \HCOP\  \jfour & 7.54 & 0.05 & 21 \\
\enddata
\end{deluxetable}


\subsection{Optical Depth Profiles}

\begin{figure*}
\epsscale{1.0}
\plotone{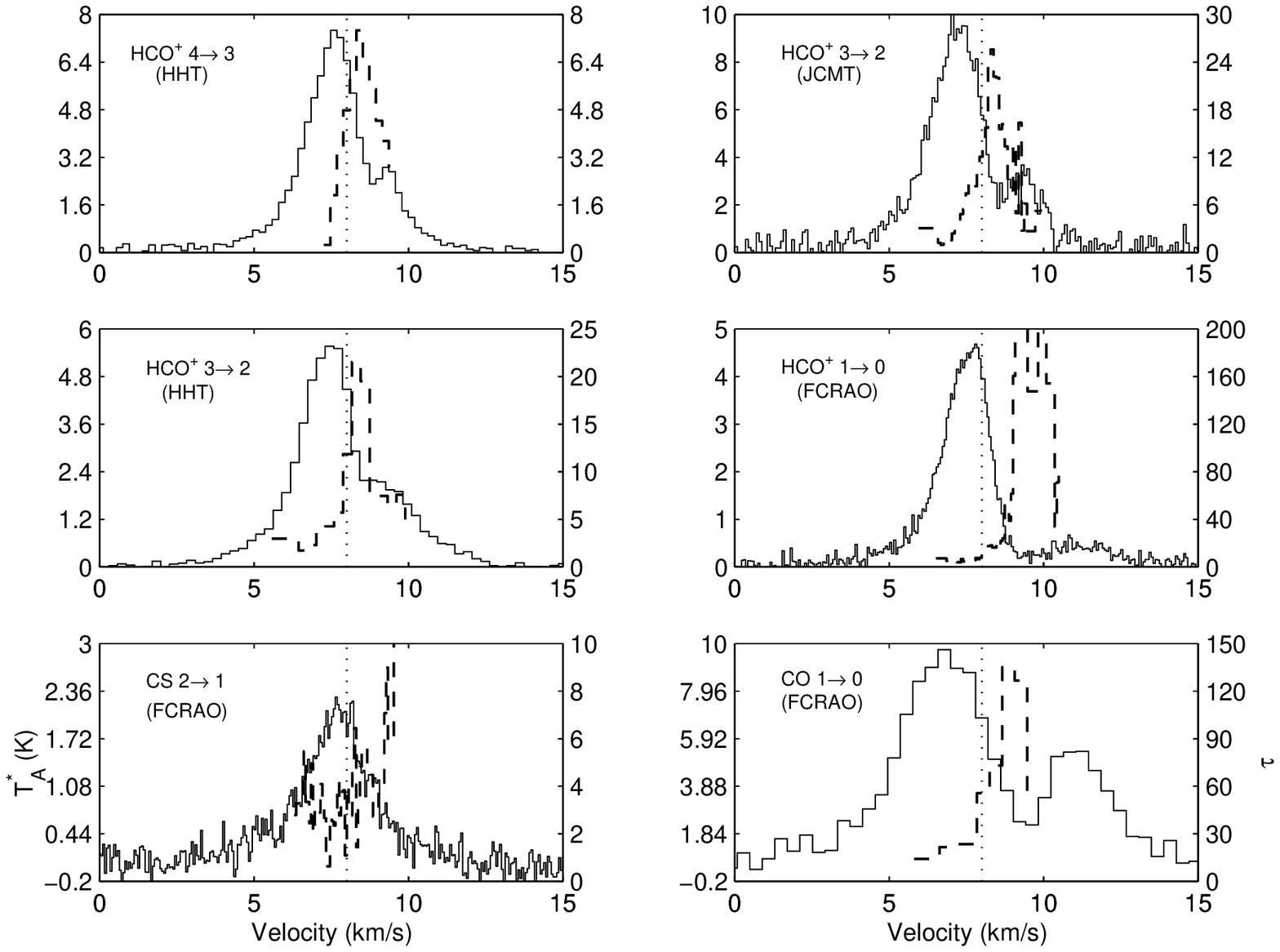}
\caption{\HCOP, CS and CO optical depth profiles towards the
central position of SMM4. For each transition, the main isotope line
profile is shown in light solid histograms, while its corresponding
optical depth profile is shown in dashed, heavy histograms. The
temperature scale is on the left-hand side Y axis, while the opacity
scale is shown on the right. The vertical dotted line represents the
systemic velocity ($8$ kms$^{-1}$) of the source. See text for details
on the computation of opacities.\label{tauall}}
\end{figure*}

\placefigure{tauall}

In the presence of infall, isotopic line emission would always be
expected to be redshifted with respect to the main line \citep{nwb98}.
In the presence of infall alone, {\em both} the emergent line
intensity and the emergent optical depth profile must be asymmetric;
the line intensity would be greater at blueshifted velocities, and
correspondingly, the line optical depth would be greater at redshifted
velocities. Indeed, the red asymmetry in the optical depth profiles of \HCOP\
\jfour\ emission derived from main and isotopic observations were used
to bolster the case for infall in the IRAS 16293-2422 Class 0 system
\citep{nwb98}. In the presence of pure expansion (outflow), the
converse would be true: the line profile would be red asymmetric, and
the optical depth profile would be blue asymmetric.  The recently
proposed quantitative indicator of infall, $\delta V$ which is the
difference in the velocity of peak emission between main and isotopic
lines, normalized by the FWHM of the isotopic line \citep{mar97}, is
based on this effect; $\delta V$ would be more negative for regions
with stronger infall velocity fields.

To determine the nature of the optical depth profiles, we performed an
optical depth analysis of the observed \HCOP, CS and CO lines. Figure
\ref{tauall} shows the resultant optical depth profiles towards the
central position of SMM4. The optical depth calculations were
performed only over the FWHM of the isotopic line. The opacities were
estimated from the observed line profile ratios of \HCOP/\HICOP,
CS/C$^{34}$S, and CO/C$^{18}$0. The opacity, $\tau_\nu^M$ of the more
abundant isotope was estimated using $I(Main)/I(Iso) =
(1-e^{\tau_\nu^M})/(1-e^{\tau_\nu^{M/r}})$, where $r$ is the ratio of
main to isotopic abundances, and $I(Main)$ and $I(Iso)$ are the line
intensities of the main and isotopic lines respectively.  The main
line is smoothed to the velocity resolution of the corresponding
isotope, and the opacity calculation is performed for all velocities
within the FWHM of the isotopic line. This technique assumes that the
excitation temperature is the same for the main and isotopic
species. Isotopic abundance ratios of [\HCOP]:[\HICOP]$=$ 65:1,
[CS]:[C$^{34}$S]$=$ 22.5:1, and [CO]:[C$^{18}$O]$=$ 280:1 are assumed
\citep{wil94}.

From Figure \ref{tauall}, it is seen that all the observed optical
depth profiles towards the central position of SMM4 shows a red
asymmetry, which is consistent with an infall scenario towards this
object. Moreover, it is seen that for similar angular and spectral
resolutions, \HCOP\ has greater optical depth, and hence, seems to
trace infall better than CS. The latter result is probably due to the
fact that the dipole moment of \HCOP\ is twice that of CS, and hence,
line and opacity profiles of \HCOP\ would be expected to show more
pronounced asymmetries.

\subsection{Integrated Intensity Maps}

\subsubsection{CO Maps}
\label{co-outflow}

\begin{figure*}

\epsscale{0.65} 
\plotone{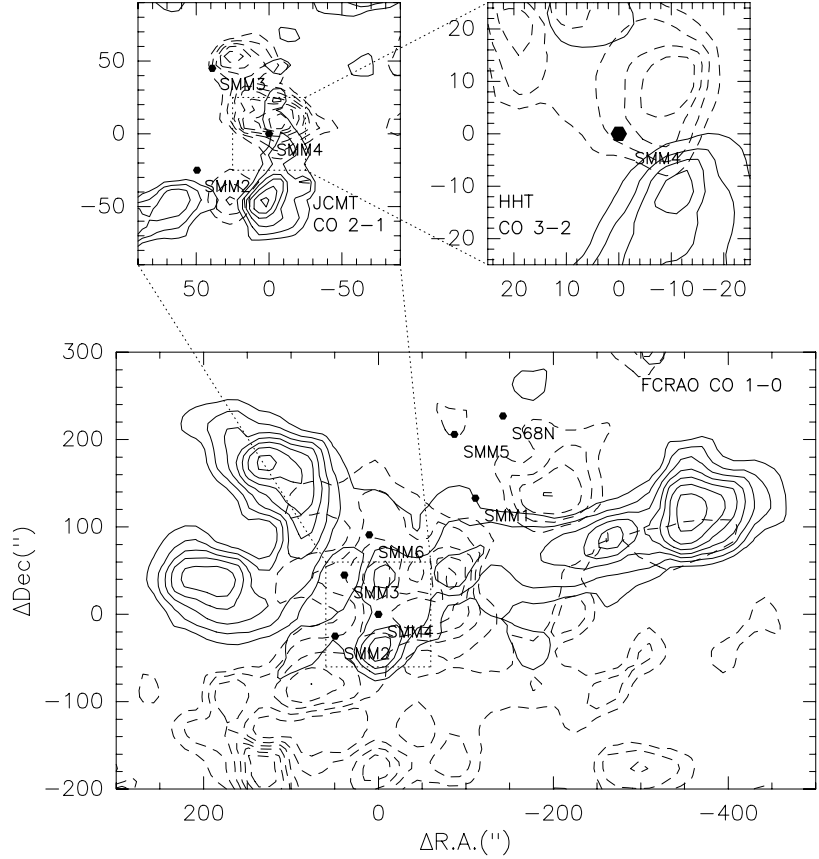}
\caption{CO Integrated Intensity Maps in line wings towards
SMM4. Three successively zoomed in views are shown. The bottom plot
shows a widefield map of extremely high velocity (EHV) wings in CO
\jone. Dashed contours show blueshifted line emission ($-10$ to $4$
kms$^{-1}$), while solid contours show redshifted line emission ($12$
to $22$ kms$^{-1}$). Contour levels are $2$ to $12.5$ by $2.5$ \Kkms,
and $4$ to $18$ by $2$ \Kkms\ for blueshifted and redshifted emission
respectively.  Submillimeter sources from \citep{cas93} are also
marked. Archival JCMT CO \jtwo\ data in the central 2\arcmin $\times$ 2\arcmin\ area
around SMM4 is shown in the top left. Dashed EHV blueshifted ($-4$ to $0$
\kms) contours from 1 to 4.5 by 0.5 \Kkms and solid EHV redshifted
($18$ to $22$ \kms) contours from 1.5 to 7 by 0.5 \Kkms are shown. To
the top right, HHT CO \jthree\ linewing emission from the central
1\arcmin $\times$ 1\arcmin\ region around SMM4 is shown. Dashed EHV
blueshifted ($-15$ to $2$ \kms) contours from 12 to 30 by 4 \Kkms, and
solid EHV redshifted ($13$ to $25$ \kms) contours from 12 to 32 by 4
\Kkms\ are shown.
\label{intco}}
\end{figure*}

The CO \jone\ and \jthree\ spectra in Figure \ref{plotall} show deep
self-absorption and line wings, suggesting the presence of outflows in
this source. Previous attempts at associating specific outflows with
the individual submillimeter sources in the Serpens cloud have shown a
complex set of flows in this region \citep{white95,wolf98,davis99}. We
performed wide-field imaging (20\arcmin $\times$ 12\arcmin) of the CO
\jone\ emission towards the Serpens region to understand the global
distribution of CO. Higher transitions of CO tend to pick up the
hotter, denser gas that may be associated with the outflow closer to
the driving source, and so we mapped a smaller region of the \jthree\
transition around SMM4. In Figure \ref{intco}, we present three
successively zoomed in views of the outflows in three different
CO transitions towards SMM4.

The EHV lobes of CO \jone~ in Figure~\ref{intco} show several
flows. At least four distinct redshifted lobes are seen: one east of
SMM3, one northeast of SMM6, one west of SMM1, and one south and west
of SMM4 and SMM2. However, corresponding blueshifted lobes that may
form a bipolar flow are seen to be less distinct. There is one
blueshifted lobe west of SMM1. In the vicinity of SMM4, blueshifted
emission occurs to the north of SMM4 and to the south and east of
SMM2. In the archival JCMT CO \jtwo\ data shown in the top left of
Figure~\ref{intco}, a distinct bipolar flow is seen in the region
around SMM4, with EHV blueshifted emission to the north-east of SMM4,
and redshifted emission to the south-west. The lobes seen in the
\jtwo\ data have good correspondence with the \jone\ data. In addition
the \jtwo\ EHV emission does not seem to trace the lower excitation
material seen in the \jone\ EHV emission. This detection of outflow
towards SMM4 is confirmed in the CO \jthree\ map obtained with the HHT
(top right plot of Figure~\ref{intco}), which also shows the same
orientation and direction of the outflow. Curiously, in the \jtwo\ map
of Figure~\ref{intco}, the blueshifted lobe of emission north-east of
the driving source has a different position angle ($25$\arcdeg$\pm
15$\arcdeg\ measured east of north) from the redshifted lobe of
emission which seems to be directed almost due south of the driving
source. A similar morphology is seen in the CO \jthree\ map, with the
p.a. of the blueshifted lobe being even further eastward
($45$\arcdeg$\pm 20$\arcdeg\ east of north) than in the \jtwo\
transition\footnote{The uncertainties in the position angles are
estimated from the angles subtended from the driving source location
to the edges of the half-power contours in the corresponding
integrated intensity map.}. Close to the driving source, it is
possible that the redshifted lobe has a similar p.a. to the
blueshifted one. But further downstream ($\sim$ 30\arcsec~ away), the
redshifted lobe becomes more north-south in direction, possibly due to
some interaction of the outflow with the ambient cloud material. The
difference in the position angles of the blue and redshifted lobes of
the outflow might be due to precession in the underlying jets. Higher
angular resolution observations will be required to test this
hypothesis.

\subsubsection{\HCOP~ Maps}
\label{hcoint}

\begin{figure*}
\epsscale{0.95}
\plotone{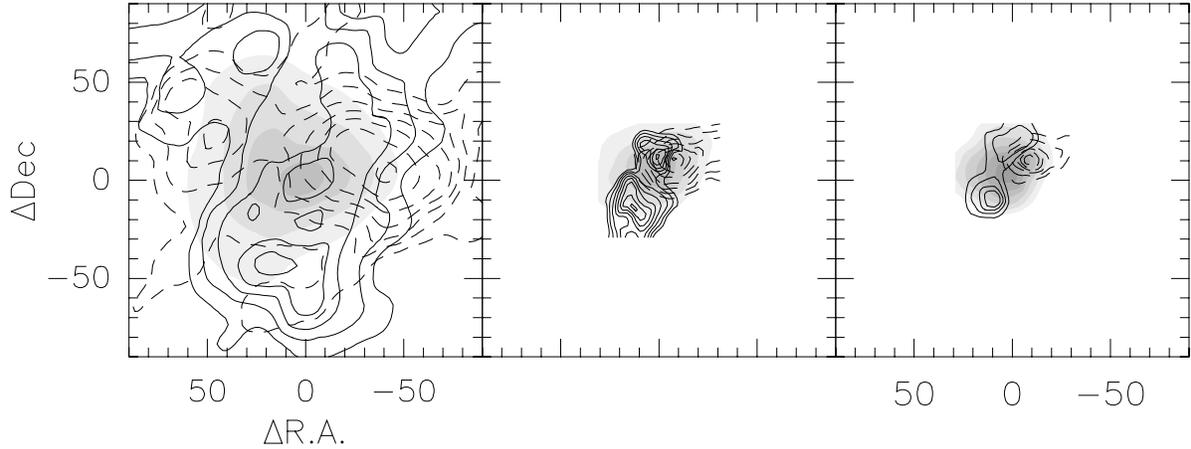}
\caption{\HCOP~ integrated intensity maps towards SMM4. In all three
panels, blueshifted emission is shown in dashed contours, while
redshifted emission is shown in solid contours. Line core emission
which is expected to mostly trace the cloud core material alone is
shown in gray scale. Left Panel: 
FCRAO \HCOP~ \jone~ integrated intensity maps. Blueshifted wing
emission ($-1$ to $6$ \kms), linecore (6 to 10 \kms) and redshifted
wing emission (10 to 16 \kms) are shown in contour levels of 0.6 to
1.8 by 0.2, 2 to 9 by 1, and 0.7 to 1.8 by 0.2 \Kkms~
respectively. 
Middle Panel: JCMT \HCOP~ \jthree~ integrated intensity
maps. Blueshifted wing
emission ($0$ to $6$\kms), linecore (6 to 10 \kms) and redshifted
wing emission (10 to 15 \kms) are shown in contour levels of 2 to
8 by 0.75, 10 to 22 by 3, and 1.5 to 3 by 0.2 \Kkms~
respectively. Right Panel: HHT \HCOP~ \jfour~ integrated intensity
maps. Blueshifted wing
emission ($0$ to $6$ \kms), linecore (6 to 10 \kms) and redshifted
wing emission (10 to 16 \kms) are shown in contour levels of 1.5 to
3.2 by 0.3, 7 to 13 by 1, and 1 to 1.4 by 0.1 \Kkms~
respectively. 
\label{inthco}}
\end{figure*}

The blue-bulge signature is expected to be seen in the centroid
velocity maps of optically thick, density sensitive tracers like
\HCOP, which preferentially traces the infalling dense cloud core. On
the other hand if \HCOP~ emission arises mostly from the outflow
material, the interpretation of the centroid velocity signature
becomes less clear. Do the molecular outflows mapped in CO have an
effect on the \HCOP\ line profiles? To determine this, we made
integrated intensity maps of the blueshifted, linecore and redshifted
emission of HCO+ \jone, \jthree, and \jfour. Figure~\ref{inthco} shows
the maps, where the integrated intensity in the line core is shown in
grayscale, while the blueshifted and redshifted integrated intensity
maps are shown in dashed and solid line contours respectively. While
the linecore integrated intensity maps are expected to trace the dense
(infalling) molecular cloud core, the linewing maps might indicate if
the \HCOP~ emission is arising from material entrained by the outflow.

The \HCOP~ \jone~ linecore integrated intensity map of
Figure~\ref{inthco} is centered on SMM4, but has a morphology that is
extended along the direction of the CO outflow, as well as orthogonal
to it. The elongation of the linecore emission orthogonal to the
outflow might be due to the embedded dense cloud core. At the same
time, the component of the \jone~ linecore emission that is elongated
along the outflow direction might be taken as morphological evidence
for the contamination of \HCOP~ \jone~ linecore emission from material
associated with the high velocity molecular outflow.  The \jone~
linewing emission also follows these orthogonal directions, but shows
heavy contamination by the outflow. The peak emission of the
redshifted outflow lobe seen in the JCMT CO \jtwo\ map (see Figure
\ref{intco}) at an offset of ($10$,$-50$) is also seen in the \HCOP~
\jone~ redshifted linewing emission, indicating that \HCOP~ \jone~
linewing emission is probably arising from the swept-up dense shells
of the underlying outflow.

\begin{figure*}
\epsscale{0.65} 
\plotone{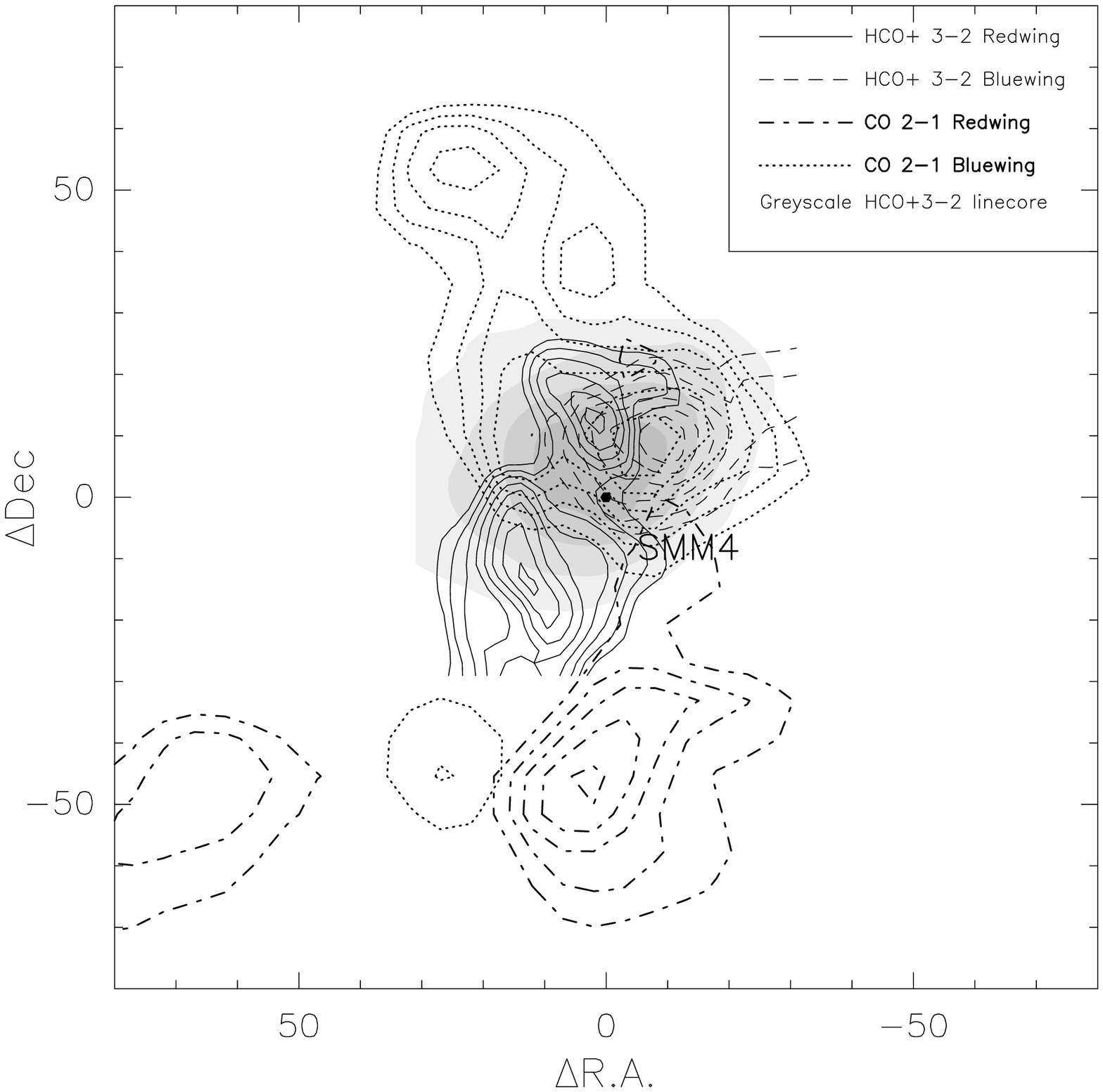}
\caption{Outflow Lobes traced by JCMT CO \jtwo\ in line wings
superposed over \HCOP\ \jthree~ emission. The outflow map from
Figure~\ref{intco}b is shown superposed over the \HCOP~
emission from Figure~\ref{inthco}b. 
\label{cohco}}
\end{figure*}

Higher-lying transitions of \HCOP~ have higher critical density and
excitation requirements and are expected to be less contaminated by
outflow. Indeed, the integrated intensity maps of JCMT \HCOP~ \jthree~
emission in Figure~\ref{inthco} show very little evidence for
extension along the outflow. The linecore map appears elliptical and
is centered on SMM4, and probably traces the infalling, dense cloud
core. The morphology of the linewing emission is a little more
complicated. There are two distinct lobes of redshifted emission. The
one in the north is probably associated with the redshifted component
also seen to the north in the CO map (see Figure~\ref{intco}). The
northern redshifted lobe might be associated with the SMM3
source. Other than this, there is a blueshifted lobe to the northwest
and a redshifted lobe to the southeast. 
To study in detail the effects of the outflow on the \HCOP~ line
profiles, in Figure~\ref{cohco}, we superpose the JCMT CO \jtwo~
outflow maps of Figure~\ref{intco}b with the JCMT \HCOP~ \jthree~ maps
from Figure~\ref{inthco}b.  Several important conclusions can be drawn
from this figure. The \HCOP\ \jthree~ linewing and linecore emission
have similar position angles, and it is approximately orthogonal to
the direction of the outflow, which indicates that both linecore and
linewing submillimeter \HCOP\ emission might be tracing the dense
starforming cloud core. The redshifted lobe of \HCOP\ \jthree~
emission is clearly anticorrelated with any outflow emission. On the
other hand, the blueshifted linewing emission of the \HCOP~ \jthree~
transition is positionally coincident with the blueshifted lobe of the
CO outflow. It is possible that the blueshifed outflowing gas is
interacting with dense ambient gas in the foreground giving rise to
the blueshifted \HCOP\ emission seen northwest of SMM4. Due to
possibility that some portion of the linewing emission of \HCOP~ might
be contaminated by outflow, in the next section, when we derive the
centroid velocity maps, we will only use the linecore of \HCOP\
emission to trace any infall signatures.  The {\em linecore} emission
of \HCOP\ \jthree~ is flattened and elliptical and appears orthogonal
to the outflow.  From the elliptical lobe of emission in the \HCOP\
\jthree~ linecore (see Figures \ref{inthco} and \ref{cohco}), we
estimate a position angle of $60$\arcdeg$\pm 10$\arcdeg\ west of north
for the major axis of the ellipse. If this elliptical lobe of emission
traced in the \HCOP\ \jthree~ linecore is tracing a rotating cloud
core, then it could be used to constrain the position angle of the
outflow, as the outflow would be expected to be orthogonal to the
orientation of the cloud core. The approximately
northeast-to-southwest direction implied by the blueshifted outflow
lobe is more consistent with the orientation of the \HCOP\ cloud core,
than a north-south outflow implied by the redshifted lobe of the
outflow (see Figure~\ref{intco}).  Our numerical modeling of the
infall region (see \S\ref{infall-model} below) is also more consistent
with a $\sim 45$\arcdeg~ position angle than a north-south orientation
for the rotational axis.  A rotating type of structure was also seen
in the case of the protobinary IRAS 16293, where submillimeter
linecore and linewing emission of CS \jseven~ and \HCOP~ \jfour~
showed evidence for a rotating, circumbinary structure \citep{nwb98}.

The \HCOP~ \jfour~ maps seen in the rightmost panel of
Figure~\ref{inthco} also show a similar result. Here too, the lincore
emission is elongated orthogonal to the outflow, and the \HCOP~
linewings seem to trace rotation in the cloud core. The orthogonality
of the \HCOP~ \jfour~ linewing emission from the EHV linewing emission
of CO \jthree~ (see Figure \ref{intco}) is even more pronounced. It is
also seen that the peaks of redshifted and blueshifted emission of
this next higher transition are located closer to the center of SMM4
compared to the \HCOP~ \jthree~ lobes, probably due to the enhanced
excitation conditions closer to the center of the object.

In summary, the linewing emission of \HCOP\ becomes progressively less
associated with outflow as we go up the rotational ladder. The
linecore emission of the \jthree~ and \jfour~ transitions of \HCOP~
are likely not impacted by the outflow, while the \HCOP~ \jone\
linecore emission might be tracing both the embedded cloud core as
well as dense shells of the molecular outflow.

\subsection{Centroid Velocity Maps}

The centroid velocity of a line profile is that velocity at which the
integrated intensity (the area under the line profiles) is equal on
either side. Centroid velocity maps have been shown to be a better
tool in the detailed study of complicated velocity fields than
integrated intensity maps \citep{ade88,nwb98}. Centroid velocity maps
of CS and \HCOP\ emission derived from numerical models of collapse
have been shown to be a good indicator of underlying infall {\em
and} rotational velocity fields (\citet{nar98}, hereafter NW). In
the model isovelocity maps of NW, the rotational velocity field
imposes a gradient of blueshifted to redshifted velocities, with the
sense of the gradient being orthogonal to the rotational axis. When
infall dominates in the central regions, the line profiles in the
central region become blue asymmetric, and hence there is a
preponderance of blue-shifted velocities in the central regions of the
isovelocity maps, giving rise to the so-called ``blue-bulge''
signature.  NW also showed that if the angular momentum content in a
given cloud is low, or if the infalling cloud core is observed with
low angular resolution, then the blue-bulge signature is
``washed-out'', that is the central regions of the isovelocity maps
just consist of blue-shifted velocities. The first detection of the
predicted blue-bulge signature was reported towards IRAS 16293
\citep{nwb98}. 

\begin{figure*}
\epsscale{0.95}
\plotone{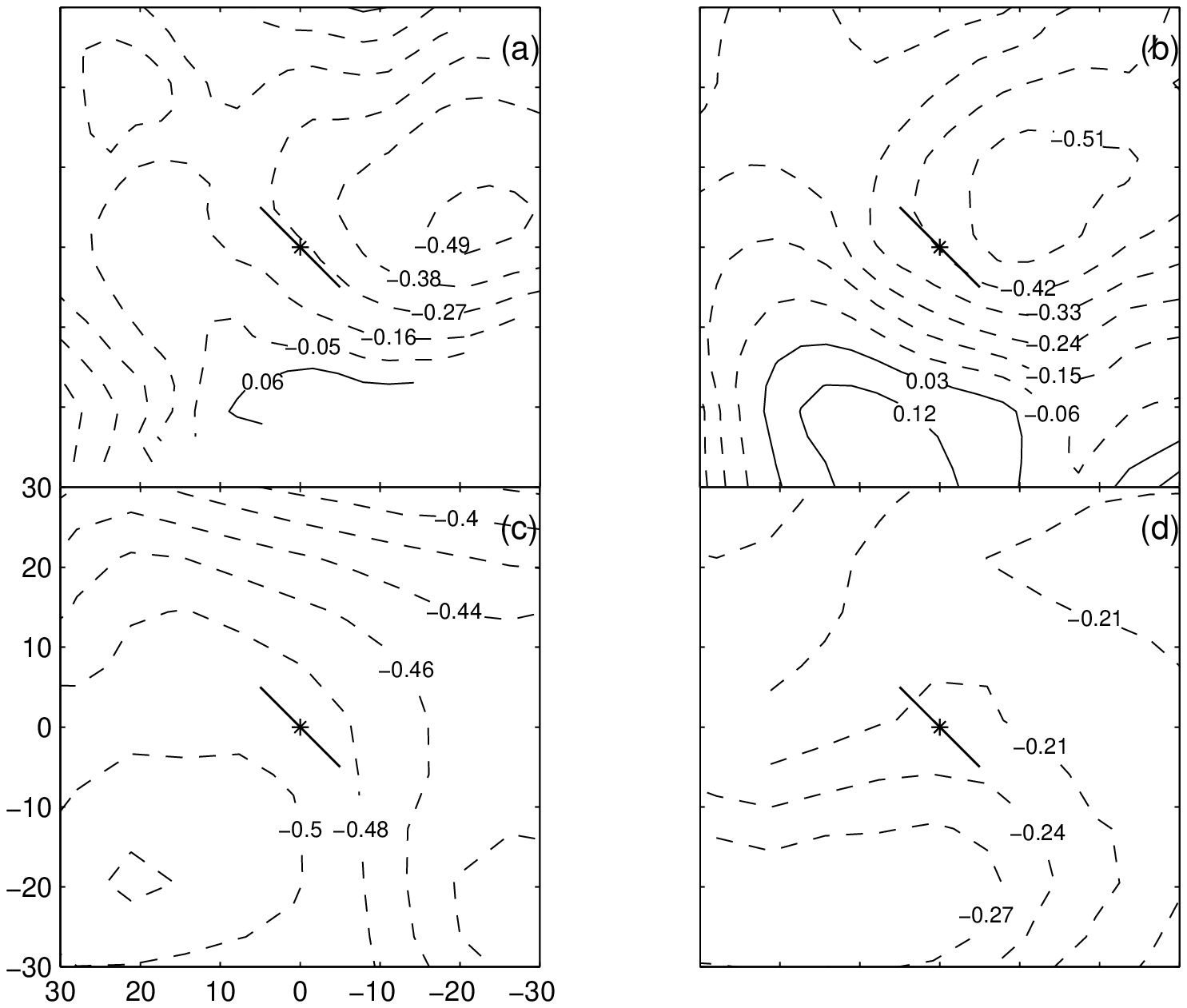}
\caption{Centroid Velocity Maps of \HCOP and CS towards SMM4.  The
centroid velocities are obtained over the velocities corresponding to
the line core (6 to 10 kms$^{-1}$). The v$_{LSR}$ (8 kms$^{-1}$) of
the source has been subtracted in the maps. Blueshifted velocities are
shown in {\em dashed} contours, and redshifted velocities are shown in
{\em solid} contours. The contour levels are labeled on the
contours. The central 60\arcsec\ $\times$ 60 \arcsec\ is shown.  Only
the observed points where the integrated intensity is greater than
five times its corresponding rms uncertainty is used in making the
centroid velocity map. The approximate location of the driving source
and the direction of the outflow are marked in each map with a star
and line respectively.  (a) HHT \HCOP\ \jfour, (b) JCMT \HCOP\
\jthree, (c) FCRAO \HCOP\ \jone, and (d) FCRAO CS
\jtwo.\label{centall}}
\end{figure*}

In Figure \ref{centall}, we present the centroid velocity maps of the
observed \HCOP and CS transitions towards SMM4. The centroid
velocities are expressed with respect to the v$_{LSR}$ of the source,
i.e., we have subtracted the v$_{LSR}$ (8 kms$^{-1}$) from the
centroid velocities, making negative velocities (shown with dashed
lines) blueshifted, and positive velocities (shown with solid lines)
redshifted. In an effort to separate out the effect of outflow
velocity fields from the dynamics of the cloud core, the velocity
centroids were computed over {\it line core} velocities. As shown in
\S\ref{hcoint}, \HCOP~ linecore emission, especially of the
submillimeter transitions, is mostly distinct from effects of
molecular outflow.

All four maps shown in Figure \ref{centall} show an increasing
gradient in centroid velocity from the northwest to the southeast. The
contours are seen to be more or less parallel, and is approximately
orthogonal to the major axis of the ellipse of \HCOP\ linecore
emission. The gradient is more ordered and pronounced in the
submillimeter transitions. The blueshifted emission in all four
centroid velocity maps are seen to encroach well south of the
rotational axis and the central source at (0,0), pushing redshifted
velocities to the extreme southeastern edge of the map. In
Figure~\ref{centall}, the submillimeter maps of \HCOP~\jthree\ and
\jfour, probably because of the higher angular resolution (19\arcsec~
and 21\arcsec~ respectively), show the blue-bulge signature more
clearly.
The millimeter maps of CS and \HCOP~ (bottom panels of
Figure~\ref{centall}) are at lower angular resolution (55\arcsec~ and
60\arcsec~ respectively), and as predicted by theory \citep{nar98}, at
lower angular resolution the blue-bulge signature appears washed out,
so that only blueshifted velocities are seen. As was discussed above,
even the linecore emission of millimeter \HCOP~ is affected by
outflow. However, the gradient seen in the bottom panels of
Figure~\ref{centall} is not north to south or north-east to
south-west, as might be expected if the millimeter lincore emission
was only purely tracing the outflow. The gradient is
northwest-southeast which is approximately orthogonal to the outflow.
This argues that the dominant underlying velocities traced by even the
millimeter CS and \HCOP~ linecore emission are rotation and
infall. Thus, all four centroid velocity maps in Figure~\ref{centall}
are consistent with the detection of the blue-bulge infall signature
towards SMM4.

\section{Discussion}
\subsection{Infall Parameters Using TSC Models}
\label{infall-model}
 
One of the thorniest problems in the unambiguous detection of infall
is the kinematic confusion that exists in protostellar regions. The
velocity fields associated with protostars include {\it infall,
rotation, turbulence and outflow} motions. To disentangle these
motions, one needs an accurate physical model for the protostar and a
flexible radiative transfer code to predict the appearance of emergent
spectra. Such model calculations will help us derive {\it
observational} diagnostics and procedures to disentangle the
kinematics in {\it real} protostellar systems. In addition, such
models can be used to constrain observations and derive useful
physical information of the protostellar region under study.

\begin{figure*}
\epsscale{0.95}
\plotone{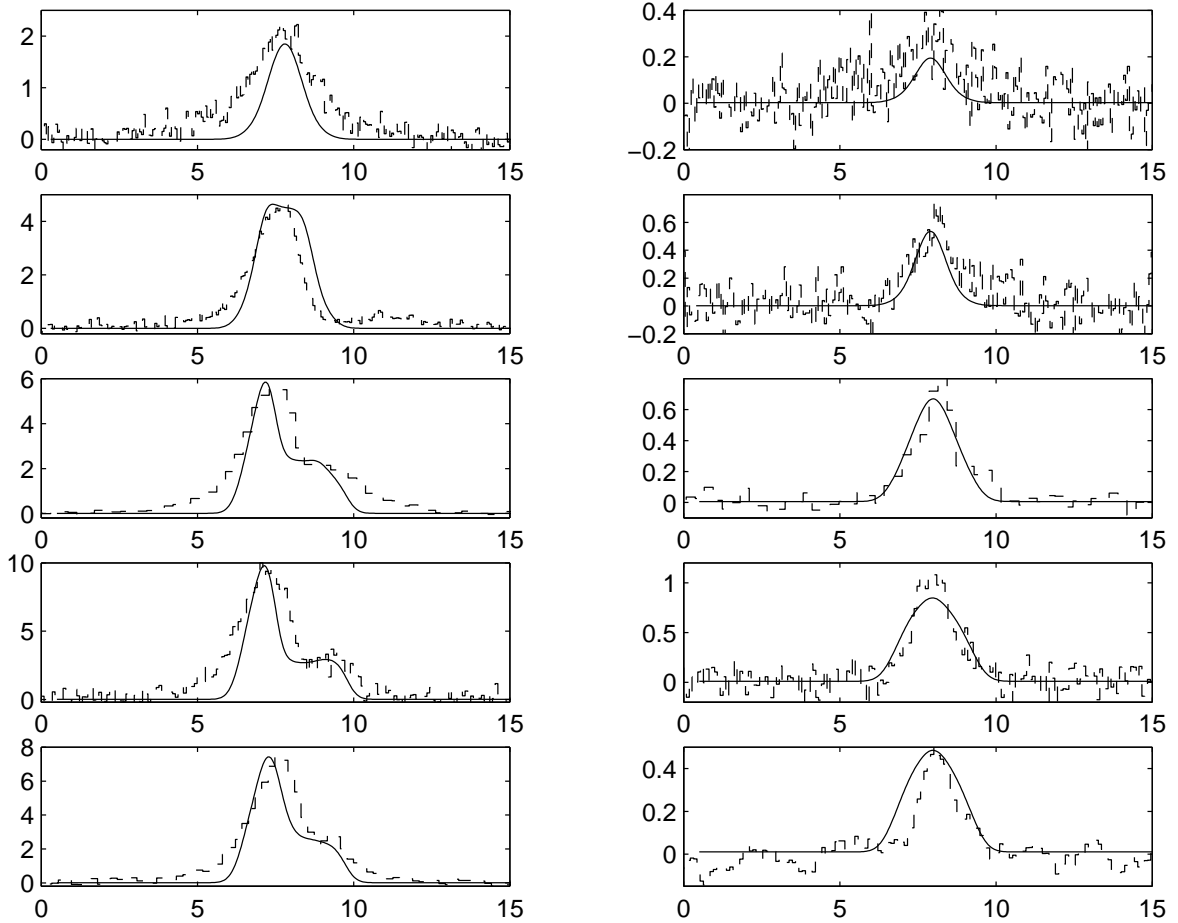}
\caption{Model Fits Towards the Central Spectra of SMM4.  Observed
spectra are shown in histograms, while the best-fit (see
Table~\ref{modelfits}) model spectra are shown in solid lines.  The
main isotope is shown in the left column, and the corresponding rarer
isotopic transition on the right column. The main transitions from top
to bottom are CS \jtwo, \HCOP~\jone, HHT \HCOP~\jthree, JCMT
\HCOP~\jthree, and HHT \HCOP~\jfour, while the corresponding isotopic
transitions on the right are C$^{34}$S \jtwo, H$^{13}$CO$^+$ \jone, 
HHT H$^{13}$CO$^+$ \jthree, JCMT H$^{13}$CO$^+$ \jthree, and HHT
H$^{13}$CO$^+$ \jfour. 
\label{modelcomp}}
\end{figure*}

While there have been several studies of line formation in a
spherically symmetric cloud core undergoing only infall motions
\citep[e.g.][]{wal86,zho92,cho95}, only a few studies have included
rotation into the mix \citep[e.g.][]{zho95,wnb94,nwb98,nar98}.  We
follow the prescription outlined in \citet{nwb98} in using the TSC
\citep{tsc84} solution of protostellar collapse to model our
observations. The TSC solution naturally accounts for infall and
rotation. In Table~\ref{modelfits}, we present the results of fitting
all our observations with the 3-dimensional TSC infall models. The
infall radius r$_{inf}$, sound speed $a$, the time since onset of
infall t$_{inf}$ (t$_{inf} = \rm{r}_{inf}/a$), rotational rate
$\Omega$, and the turbulent velocity used in the radiative transfer
solution are summarized in Table~\ref{modelfits}.  About fifty model
runs were made by varying the infall parameters each time. The
synthetic observations were then convolved to the resolution of the
telescopes used in the observations presented in this paper. For the
modeling, the main and isotopic lines of all the observed CS and
\HCOP~ data from this work were used to obtain best fits to the
observations. In the fitting process, the entire mapping data obtained
observationally are compared with the synthetic model maps. The best
fit model is obtained ``by eye'', comparing not only the spectral
profiles, but also the integrated and centroid velocity maps. While
efforts were made to ensure goodness of fit for the entire line
profile, more emphasis was laid on the linecore (which is expected to
be less impacted by outflow). Figure~\ref{modelcomp} shows the
best-fit model spectra overlaid on the observed spectra for the main
and isotopic lines towards the central position of SMM4. While only
the central spectra are shown for each transition, the fits are good
through most of the mapped regions.

\placetable{modelfits}

\begin{deluxetable}{lc}
\tablecaption{Best Fit Model Results . \label{modelfits}}
\tablewidth{0pt}
\tablehead{
\colhead{Parameter} & \colhead{Value}}
\startdata
$a$ & 0.5 \kms \\
$\Omega$ & $7\times 10^{-14} \ s^{-1}$ \\
r$_{inf}$ & 0.015 pc\\
t$_{inf}$ & $3\times 10^4$ yrs \\
v$_{turb}$ & 0.75 \kms\\
\enddata
\end{deluxetable}

The TSC model is a parameterized model, and hence the search space to
obtain the best fit model is quite large.  For this reason,
observational constraints when available, were used to set the initial
value of the infall parameters in the fitting process. The initial
value of the infall radius r$_{inf}$ was constrained using the ratio
of \HCOP~ \jthree~ main beam temperatures towards the central source
obtained from JCMT and HHT. As shown in \citet{nwb98}, it is possible
to constrain the emitting source size from the ratio of the main beam
temperatures of the same transition obtained at two different
telescopes with different beam sizes. At 267 GHz, the main beam
efficiencies are 0.6 and 0.71, and beam sizes are 19\arcsec~ and
28\arcsec~ respectively for JCMT and HHT, which can be used to obtain
the emitting source size of SMM4 to be 10.5\arcsec. This size is in
very good agreement with the core size derived from molecular emission
observed with high angular resolution aperture synthesis observations
\citep{hog99}.  Now that r$_{inf}$ is constrained, if the time since
onset of collapse t$_{inf}$ is known, the initial value of the
effective sound speed $a$ can be determined. It is now believed that
the pure infall phase before the onset of molecular outflows is
extremely short-lived or non-existent \citep{and93,sar96}. Hence the
dynamical time-scale of the observed outflows can be used as a good
indicator of t$_{inf}$. From the projected EHV lobes in the JCMT CO
\jtwo~ data shown in Figure~\ref{intco} and the maximum flow velocity,
we derive an outflow dynamical timescale of $\sim 10^4$ years, which
was used as the initial value of t$_{inf}$ in the model fits. The
angles subtended by the rotational axis in and out of the plane of the
sky are also input parameters to the models. The angle in the plane of
the sky is taken to be $45\arcdeg$ (east of north) from the appearance
of the outflow lobes in Figure~\ref{intco}. We initially assumed that
the outflow was in the plane of the sky, and the best-fit models are
consistent with this interpretation. The best value of v$_{turb}$ was
chosen from the linewidth of our H$^{13}$CO$^+$ observations. For our
model fits we used a temperature distribution profile of
$T=10(r/0.02\rm{~pc})^{-1.0}$ K. In the radiative transfer program we
used a \HCOP to H$_2$ abundance of $2\times 10^{-9}$ and CS to H$_2$
abundance of $1\times 10^{-9}$. For isotopic ratios, we used
[$^{12}$C/$^{13}$C] $= 45$ and [$^{32}$S/$^{34}$S] $=15$.

Since the fitting is done ``by eye'', it is difficult to claim that
the best fit model is a unique fit to the observed data. While some
model parameters are only loosely constrained, the modeling exercise
does set strong constraints on other parameters. For example, the
direction and size of the gradient seen in the centroid velocity maps
shown in Figure~\ref{centall} can only be reproduced in the models, by
assuming a rotational axis at an $\sim 45$\arcdeg~ position
angle. Assuming a north-south or east-west orientation of the
rotational axis results in model centroid velocity maps that have
gradients west to east and north to south, not the north-west to
south-east sense seen in Figure~\ref{centall}. The angle that the
rotational axis makes to the plane of the sky is also well
constrained. Significant departures from the assumed orientation,
viz., the rotational axis is in the plane of the sky, results in bad
fits to the ratio of blue-to-red peaks of the \HCOP~ lines. 

\subsection{Infall Size Scales, Velocities and Rates}

The best fit infall size of $0.015$ pc corresponds to an angular
diameter of $\sim 20$\arcsec~ at the distance of the Serpens
cloud. The \HCOP~ estimated core size (as discussed above) is
10\arcsec, so this indicates that the expansion wave might have
already escaped the densest regions of the core. As seen in
Figure~\ref{intco}, other protostellar sources are located quite close
to SMM4. The TSC model is applicable only to isolated collapse
resulting in a single protostar. The values derived above for the
infall parameters of SMM4 using the TSC model should thus be
interpreted with some caution. If we assume that the constrained
infall parameters are not heavily affected by the cluster environment,
we can quantify some other physical parameters in the SMM4 system.

From Table~\ref{modelfits}, the infall timescale is given by $t =
r_{inf}/a = 3\times 10^4$ yrs. This number is in good agreement with
the lower limit to the age of the outflow system derived as $\sim
10^4$ yrs in this work (see above), as well as the estimate of
$3\times 10^4$ years from the CO observations of \citet{davis99}. This
would indicate that outflow was triggered almost simultaneously with
the onset of collapse in SMM4. The rotational velocity, $\Omega =
7\times 10^{-14}$ s$^{-1}$ implies that the turnover radius $r_c =
a/\Omega = 0.23$ pc $= 150\arcsec$. From the self-similar solution of
Shu (1977), a mass accretion rate of $0.975a^3/G = 2.9\times 10^{-5}$
\Msun yr$^{-1}$ is derived. The present mass of the star-disk system
can then be calculated by $\dot{M}t$ giving a mass of $\sim 0.9$
\Msun, which falls well within the dust mass limits derived by
\citet{hog99} for SMM4. The centrifugal radius (or the accretion disk
radius) of the SMM4 system at the present epoch is $R_c =
\Omega^2G^3M^3/16a^8 \approx 0.9$ AU.

It is expected that the observed luminosity of $\sim 15$ \Lsun\
\citep{hog99} in SMM4 is probably dominated by the process of
accretion. If we assume all the observed luminosity arises from
accretion onto one star, we can estimate the mass of the star as
follows: the bolometric luminosity is given by
$L_B=\frac{G\dot{M}M_*}{R_*}$, where R$_* \approx 3$ \Rsun, and
$\dot{M} \sim 2.9 \times 10^{-5}$ \Msun yr$^{-1}$. This gives $M_*\sim
0.3$~\Msun. Since the mass of the star-disk system is $\sim 0.9$
\Msun, the mass of the central disk is $\sim 0.6$ \Msun. The maximum
value of the disk mass that is stable to m$=1$ gravitational
instability is given by $\frac{M_D}{M_D+M_*} \le 0.24$
\citep{ada93}. In the case of SMM4, this ratio is 0.67, which implies
that the disk should be gravitationally unstable against collapse or
fragmentation. From the above discussion, it appears that both the
cloud core and accretion disk (if present) are in a state of dynamical
collapse.

\subsection{Robustness of the Infall Interpretation}

The observational identification of dynamical collapse in the early
stages of star formation has been a subject of repeated controversy
\citep[see][for a review]{zho94}. Given the history of the
subject, it is natural to worry about whether our identification of
collapse in our sources is indeed the only unique conclusion. Below we
discuss some alternate models and methods of their elimination.

In identifying the classic asymmetric blue peak signature of
an observed line profile with infall, we have to worry about four
possible alternatives: (1) two cloud components along our line of
sight, and the blueshifted cloud happens to be stronger; (2) a
background component that is being absorbed by an unassociated
foreground component that happens to be redshifted; (3) an outflow
source with a stronger blue lobe; (4) the blueshifted part of a
rotating cloud. Statistically speaking, all four models listed above
have a 50\% chance of producing the asymmetric blue profile. The
infall scenario has a 100\% chance of producing the asymmetric blue
signature. In observing a large number of infall candidates, the preponderance
of blueshifted asymmetry in the central line profiles can be
statistically interpreted as identification of infall
\citep{greg97}. However, for individual sources, the blue asymmetric
line profile alone is not unique evidence for infall. 

If we observe a second line that is optically thin, we can start
eliminating some of the models mentioned above. In model (1), the
optically thin line would have two peaks or a single peak aligned in
velocity with one of the two cloud components. In model (2), the
optically thin line should show a single peak aligned in velocity with
the background component. On the other hand, if the observed optically
thin line has a single peak in the absorption dip of main line, this
would eliminate both models (1) and (2). Another effect to study when
observing an isotopic line is to use line ratios of main to isotopic
transitions, as was done in this work, to derive an optical depth
profile. Infall would be expected to produce an asymmetric red profile
in optical depth, in addition to an asymmetric blue peak in the line
profile. In models (3) and (4) we would not expect the red asymmetry
a-priori in the optical depth profile. For instance, a strong
blueshifted outflow lobe (model 3), in the presence of an increasing
temperature gradient toward the center of the source, would be
expected to have an asymmetric blue optical depth profile as well as
an asymmetric blue line profile. However, to eliminate models (3) and
(4) convincingly, we need line maps to determine the effect of
outflows and rotation.

Effects of outflows are not expected to be seen orthogonal to the
outflow. For sources, where outflow is in the plane of the sky, its
effects are expected to be somewhat reduced. One way to get around the
problem of outflow contamination is to choose molecular species that
are not very abundant in outflows. We have adopted this technique in
our study by choosing the \HCOP \jthree~ and \jfour~ transitions,
which because of their higher critical densities are not expected to
be excited in outflows. Detailed study of maps of the \HCOP~ transition
along with high velocity CO maps from outflows, as was done in this
work, will help us understand the effect of outflows on the
identification of the infall signature.

The blue-bulge signature obtained by mapping an optically thick
transition naturally takes into account the effect of rotation.  Are
there other interpretations other than infall when we observe the
morphology of the blue-bulge signature? Outflows can produce a
gradient from blueshifted to redshifted velocities
\citep[e.g.][]{nar96}. However, such a gradient should be in the same
direction as the outflow. The choice of the v$_{LSR}$ of the cloud is
somewhat critical in determining the distribution of blueshifted and
redshifted velocities in the centroid velocity maps. By obtaining
careful observations of an optically thin tracer (which would show a
gaussian profile about the v$_{LSR}$), such systematic errors can be
reduced. To zeroth order, however, we have found from models and
observations that the overall morphology of the gradient and bowing of
the velocity contours towards the center of the source is independent
of the choice of both the velocity window and the cloud velocity. When
the velocity gradient is orthogonal to the outflow, and the
blue-bulge morphology is seen, the simplest explanation is one
that involves infall.

SMM4 is a prototypical infall candidate in that all three conditions,
viz. the right sense of asymmetry in line profiles (blue asymmetric)
and optical depth profiles (red asymmetric), and the blue-bulge
signature in centroid velocity maps of density sensitive molecules are
seen. In summary, convincing evidence for infall in {\em individual}
sources requires both mapping and obtaining isotopic transitions. In
the case of SMM4, the mapping data and isotopic data from density
sensitive molecules are consistent with an infall interpretation.

\section{Summary of Results}

The region around the Class 0 protostellar system SMM4 in the Serpens
cloud core has been mapped in several millimeter and submillimeter
high density molecular tracers, providing the detection of kinematic
infall signatures towards this object. These observations show both
the classic blue-asymmetric line profile signature and the blue-bulge
centroid velocity signature of infall. The main conclusions of this
paper are summarized below. 

\begin{enumerate}
\item The line profiles of optically thick main isotopes of
\HCOP~\jone, \jthree~ and \jfour, and CS \jtwo~ exhibit the
characteristic blue asymmetric line profile signature consistent with
infall. Blue asymmetric line profiles are also seen in the CO \jone,
\jtwo~ and \jthree~ transitions. In all cases, the centroid velocity of
the optically thick main isotope is bluer than the optically thin
rarer isotope.

\item The \HCOP, CS and CO optical depth profiles obtained
from main and isotopic transitions show a red asymmetry
also consistent with an infall interpretation. 

\item Wide-field mapping of CO \jone~ emission shows that the Serpens
region has many criss-crossing outflows. From high signal-to-noise
ratio smaller scale CO \jtwo~ and \jthree~ maps, a bipolar outflow
associated with SMM4 has been identified.

\item Centroid velocity maps made with linecore emission of CS \jtwo,
and \HCOP~ \jone, \jthree~ and \jfour~ transitions showed the
blue-bulge infall signature. The blue-bulge infall signature is a
robust indicator of collapse in protostellar objects where infall
dominates over rotation, and one can disentangle the outflow field.

\item We used three-dimensional radiative transfer models based on the
rotating, collapse model of Terebey, Shu and Cassen and derived infall
parameters for SMM4. The constraints for the infall size is $\sim
10\arcsec\sim 3000$ AU. The dynamical age of the outflow and infall
phase of the SMM4 system is $\sim 3\times 10^4$ yrs, which falls well
within the predicted duration of the Class 0 stage of early stellar
evolution. 

\end{enumerate}

\acknowledgements

We gratefully acknowledge the staff of the HHT and the JCMT for their
excellent support during observations. Craig Kulesa and Aimee
Hungerford are thanked for performing some of the HHT observations.
Research at the FCRAO is funded in part by the National Science
Foundation under grant AST 97-25951.

\end{document}